\documentstyle[prl,aps]{revtex}
\input epsf
\input psfig

   \newcommand{\be}{\begin{equation}}
   \newcommand{\ee}{\end{equation}}
   \newcommand{\bea}{\begin{eqnarray}}
   \newcommand{\eea}{\end{eqnarray}}
   \newcommand{\upar}{\uparrow}
   \newcommand{\dn}{\downarrow}
   \newcommand{\ek}{\varepsilon_k}
   \newcommand{\ef}{\varepsilon_f}

\begin{document}
\draft
\twocolumn[
\widetext
\title{Supersymmetry and Theory of Heavy-Fermions}
\author{C. P\'epin and M. Lavagna*}
\address{Commissariat \`a l'Energie Atomique, \\ D\'epartement de
  Recherche Fondamentale sur la Mati\`ere Condens\'ee /SPSMS, \\
17, rue des Martyrs,
  \\ 38054 Grenoble Cedex 9, France}
\maketitle \widetext
  \leftskip 54.8pt
  \rightskip 54.8pt
  \begin{abstract}
We propose a new approach to the Kondo lattice in order to describe simultaneously the Kondo effect and the local magnetism. This approach relies on an original representation of the $S=1/2$ impurity spin in which the different degrees of freedom are represented by fermionic as well as bosonic variables.  The magnetic instability is found at $J_c/D =0.67$ in definite improvement compared to usual fermionic mean-field theories. The dynamical susceptibility 
contains an inelastic peak in addition to the standard Lindhard-type Fermi liquid contribution. The Fermi surface sum rule includes $n_c+1$ states corresponding to large Fermi surfaces.
\par
  \end{abstract}
\vspace{0.2in}
]
\narrowtext
An outstanding feature of heavy-Fermion systems is the coexistence of Fermi liquid-type excitations with local magnetism resulting from RKKY interactions among spins. For instance, the dynamic magnetic structure factor measured in CeCu$_6$ and CeRu$_2$Si$_2$ by inelastic neutron scattering (INS) indicates two contributions~\cite{aeppli,rossat}: a q-independent single-site slow component typical of the localized  excitations of Kondo-type, superimposed on a strongly q-dependent intersite fast component, reflecting the magnetic correlations due to RKKY interactions. The former is peaked around $\omega =0$ (quasi-elastic peak) while the latter is peaked around $\omega_0 \neq 0$ (inelastic peak). Another important probe is provided by the de Haas-van Alphen (dHvA) experiments. The results~\cite{lonzarich} obtained by different groups on various compounds indicate heavy effective mass. They also agree as to the existence of large Fermi surfaces in the magnetically-disordered phase. Even though the charge degrees of freedom are frozen, the localized electrons seem to contribute to the Fermi surface sum rule together with the conduction electrons.
\par
This coexistence of Fermi liquid-type excitations with low energy magnetic fluctuations is likely to stem from the nature of the screening of the localized moments in the Kondo lattice. The interesting domain is reached when $I\leq T_K \leq T_3$ ~     ($I$ is the order of the RKKY interaction, $T_K$ the Kondo temperature and $T_3$ the temperature-scale at which the number of thermal states for conduction electrons just equals the number of impurities). Then, at $T_K$ the available conduction electrons are exhausted~\cite{nozieres}  before achieving complete screening (incomplete Kondo effect) leaving residual unscreened spin degrees of freedom on the impurities.

Traditionally, the spin is described either in fermionic or bosonic representation. If the former representation, used for instance in the $1/N$ expansion of the periodic Anderson-PAM~\cite{millis} or the Kondo lattice-KLM ~\cite{auerbach} models, appears to be well adapted for the description of the Kondo effect, it is also clear that the bosonic representation lends itself better to the study of local magnetism. Quite obviously the physics of  heavy-Fermions is dominated by the duality between Kondo effect and localized moments. This constitutes our motivation to introduce a new approach to the KLM which relies on an original representation of the impurity spin $1/2$ in which the different degrees of freedom are represented by fermionic as well as bosonic variables. The former are believed to describe the Fermi liquid excitations while the latter account for the residual spin degrees of freedom.
\par
Let us list the main results obtained in this paper: (i) the Fermi surface sum rule includes $n_c+1$ states which means that the Fermi surface volume includes a contribution of one state per localized spin in agreement with dHvA experiments, (ii) the dynamical susceptibility contains an inelastic peak in addition to the standard Lindhard-type Fermi liquid contribution as observed
in INS, (iii) the instability towards long-range magnetic order appears at $J_c/D=0.67$
significantly reduced to the value $J_c/D=1.0$ obtained in purely fermionic mean-field theories. While (i) has been already obtained in previous
work, we stress on the fact that (ii) and (iii) constitutes new results established by this supersymmetric approach.
\par 
We propose the following mixed fermionic-bosonic representation of the 
spin $S=1/2$. 
The two states constituting the basis may be written: $|1/2,1/2\rangle =
(X b^{\dagger}_{\upar}+Y f^{\dagger}_{\upar})|0\rangle$ and $ |1/2,-1/2\rangle = (X b^{\dagger}_{\dn}+Y f_{\dn}^{\dagger})|0\rangle $, where $b^\dagger_\sigma$ and $f^\dagger_\sigma$ are respectively bosonic and fermionic creation operators, $X$ and $Y$ are parameters controlling the weight of both representations and $\left| 0 \right\rangle$ represents the vacuum of particles:  $b_\sigma \left| 0 \right\rangle =f_\sigma \left| 0 \right\rangle =0$.
The spin operators are then given by $ S^{\alpha} =  
\sum_{\sigma\sigma'}{b_{\sigma}^{\dagger}\tau^{\alpha}_{\sigma\sigma'}b_{\sigma'}+f^\dagger_{\sigma}\tau^{\alpha}_{\sigma\sigma'}f_{\sigma'}}
 = S^{\alpha}_{b}+S^{\alpha}_{f}$, where $\tau^\alpha$ $(\alpha = \left(+,-,z\right))$ are Pauli matrices. ${\bf S}_b$ and ${\bf S}_f$ are equivalent to Schwinger boson and Abrikosov pseudo-fermion representations of the spin respectively.

One can easily check that the representation that we propose satisfies the standard rules of $SU(2)$ algebra: $\left| 1/2, \pm 1/2 \right\rangle$ are eigenvectors of $S^2$ and $S^z$ with eigenvalues $3/4$ and $\pm 1/2$, $\left[S^+,S^- \right]=2 S^z$ and $\left[S^z,S^{\pm}\right]=\pm S^{\pm}$ provided that the following local constraint is satisfied:
$\sum_{\sigma}{b_{\sigma}^{\dagger}b_{\sigma}+f_{\sigma}^{\dagger}
f_{\sigma}}=n_{b}+n_{f}=1$,
which implies $ X^2+Y^2 =1$. In order to eliminate the unphysical states as $\left( X b^\dagger_\sigma + Y f^\dagger_{-\sigma} \right) \left| 0\right\rangle$, we need to introduce a second local constraint and take: 
$Q=\sum_{\sigma} \sigma (f_{\sigma}^{\dagger}b_{-\sigma}+b_{-\sigma}^{\dagger}f_{\sigma})=0$

Let us consider the three-dimensional KLM near half-filling ($n_c\leq 1$). The hamiltonian is:
\be
H=\sum_{k\sigma} {\ek c_{k\sigma}^\dagger c_{k\sigma}}+J \sum_{i\sigma\sigma^\prime} {(c_{i\sigma}^\dagger \tau _{\sigma\sigma^\prime} c_{i\sigma^\prime}).{\bf S_{i}}}
\ee
In the representation introduced before, the partition function can be written as the following path integral: 
\be \begin{array}{l}
Z=\int {\cal D}c_{i\sigma}{\cal D}f_{i\sigma}{\cal D}b_{i\sigma}
d \lambda_{i} d K_i
\exp  \Bigl[-
  \int_{0}^{\beta} d\tau ({\cal L}_0 + {\cal H} \nonumber \\
   +\sum_i \ef{_{i}} (n_{b_{i}}+n_{f_{i}}-1) + \sum_i K_i  Q_i) \Bigr] \ , \\  \\
\begin{array}{ll} 
\mbox{with} & {\cal L}_0=\sum_{i\sigma}(c_{i\sigma}^\dagger\partial_{\tau}c_{i\sigma}+f_{i\sigma}^{\dagger} \partial_{\tau}f_{i\sigma} +b_{i\sigma}^{\dagger}\partial_{\tau}b_{i\sigma}) \\ \\
\mbox{and} &  
{\cal H}=\sum_{k\sigma}{\ek c_{k\sigma}^{\dagger}c_{k\sigma}}  \\
& + J\sum_{i}{({\bf S}_{f_{i}}+{\bf S}_{b_{i}}).{\bf s}_{i}}  -\mu\sum_{i}{n_{c_{i}}} \ , 
         \end{array} \end{array} \ee 
where $J$($>0$) is the Kondo interaction, and two time-independent Lagrange multipliers $\ef{_{i}}$ and $K_{i}$ are introduced to enforce the local constraints.

Performing a Hubbard-Stratonovich transformation and neglecting the space and time dependence of the fields in a self-consistent saddle-point approximation, we have :
\be
\label{sadpoint} \begin{array}{l}
Z=\int{d\eta d\eta^{*}{\cal C}\left(\sigma_0, \lambda_0, \eta ,\eta*\right)Z(\eta,\eta^{*})} \\ \\
Z(\eta,\eta^{*})=\int{\cal D}c_{i\sigma}{\cal D}f_{i\sigma}{\cal D}b_{i\sigma}  \exp\Bigl[- \int_0^\beta d \tau \left( {\cal L}_0 + {\cal H}' \right) \Bigr] \ , \\ \\
\begin{array}{ll}
\mbox{with} 
& {\cal C}\left(\sigma_0, \lambda_0, \eta, \eta* \right) = \exp [\beta \left( \lambda_0 - \frac{\sigma_0^2+ \eta^* \eta}{J} -\mu\sum_{i}{n_{c_{i}}} \right)] \ ,\\
& {\cal H}' = \sum_{k\sigma}{(
f^{\dagger}_{k\sigma}c^{\dagger}_{k\sigma}b^{\dagger}_{k\sigma})
H_{0}\left(
\begin{array}{c}
f_{k\sigma} \\
c_{k\sigma} \\
b_{k\sigma}
\end{array}
\right)}  \ , \\    
& H_{0}= \left(
\begin{array}{ccc}
\ef & \sigma_{0} & 0 \\
\sigma_{0} & \ek & \eta \\
0 & \eta^{*} & \ef
\end{array}
\right) \,  \end{array} \end{array} \ee
Note the presence of a Grassmannian coupling $\eta$ between $c_{i \sigma}$ and $b_{i \sigma}$, in addition to the usual coupling $\sigma_0$ between $c_{i \sigma}$ and $f_{i \sigma}$ responsible for the Kondo effect. $H_0$ is of the
type $\left( \begin{array}{cc} a & \sigma \\ \rho & b \end{array} \right)$ in which $ a$,$b$ ($\rho$,$\sigma$) are matrices consisting of commuting (anticommuting) variables. Note the supersymmetric structure of the matrix 
$H_0$ similar to the supermatrices appearing in the theory of disordered
metals~\cite{efetov}. The second constraint is automatically satisfied in this scheme since there is no off-diagonal term mixing $\upar$ end $\dn$ states.

$H_0$ being hermitian, the matrix $U^\dagger$ transforming the original basis $\psi^\dagger = \left( f^\dagger c^\dagger b^\dagger \right)$ to the basis of eigenvectors $\Phi^\dagger = \left( \alpha^\dagger \beta^\dagger \gamma^\dagger \right)$ is unitary ($U U^\dagger=U^\dagger U=1$). $\Phi^\dagger 
= \psi^\dagger U^\dagger$ with $U^\dagger$ a supersymmetric matrix. 
$\alpha^\dagger$ and $\beta^\dagger$ are the fermionic eigenvectors whose eigenvalues, determined from $
\det\left[(a-E)-\sigma(b-E)^{-1}\rho\right]=0 $, are:
$$
E_{\mp} =
\frac{(\ek+\ef)\mp \sqrt{(\ek-\ef)^{2}+4(\sigma_{0}^{2}+\eta\eta^{*})}
}{2} \ .
$$
$\gamma^\dagger$ is the bosonic eigenvector whose eigenvalue, determined~\cite{ref3} from $\det\left[(b-E)-\rho(a-E)^{-1}\sigma\right]=0$ is $E_\gamma=\ef$. 

In the scheme we propose, $\sigma_0$ and $\lambda_0$ are slow variables that we determine by solving saddle-point equations, while $\eta$, $\eta^*$ are fast variables defined by a local approximation. As we will see,
the latter approximation incorporates part of the fluctuation effects. Indeed, performing the functional integration of~(\ref{sadpoint}) over the fermion and boson fields~\cite{efetov} yields a superdeterminant ($SDet$) form written as follows:
\be
\begin{array}{l}
Z(\eta, \eta^*) = SDet(\partial_\tau + H) \ , \\ \\
   \begin{array}{ll}
\mbox{where} &  {\displaystyle SDet(\partial_\tau + H) = \frac{Det(G^{-1} - \sigma D \rho ) }{ Det( D^{-1})} }\ , \end{array} \\
\begin{array}{lll}
G^{-1}= \partial_\tau + a  & \mbox{and} & D^{-1} = \partial_\tau + b \ .
\end{array} \end{array} 
\ee

Expanding to second order in $\eta$, $\eta^*$ allows us to define the propagator $G_{\eta \eta^*}( {\bf k}, i \omega_n) $ associated to the Grassmann variable $\eta$ and hence the closure relation for $x_0^2 = \left \langle \eta \eta^* \right\rangle$:
\be
\label{4}
{\displaystyle x_{0}^{2}=\frac{1}{\beta}\sum_{{\bf k},i\omega_{n}}G_{\eta \eta^*}( {\bf k}, i \omega_n)} \ ,
\ee 
$$ \begin{array}{ll}
\mbox{with} & {\displaystyle  G_{\eta \eta^*}( {\bf k}, i \omega_n) = \frac{J}{
\left[1- J\Pi_{cb}^{0}(\bf k,i\omega_{n})\right]} }\\
\mbox{and} & {\displaystyle \Pi_{cb}^0 = \frac{1}{\beta} \sum_{{\bf q}, i \omega_n} G_{cc} ({\bf k+q}, i \omega_n + i \omega_\nu) D( {\bf q}, i \omega_n) }\ .
\end{array}  $$

Contrary to \cite{gan} which in the case of the underscreened Kondo impurity model, assumes $x_0^2=0$ leading to a two-fluid description, the closure equation (\ref{4}) that we introduce defines a finite $x_0^2$. This parameter $x_0^2$ plays a major role in controlling the relative weights of fermion and boson statistics. It is directly connected to the X and Y
parameters introduced in the initial representation of the states : $X^2=x_0^2/(\sigma_0^2+x_0^2)$ and $Y^2=\sigma_0^2/(\sigma_0^2+x_0^2)$.

The resolution of the saddle-point equations, keeping the number of particles conserved, leads to:
\be \label{5} \begin{array}{c}
{\displaystyle y_{F}= -D \exp \left[ -1/(2J\rho_{0}) \right] } \ , \\
{\displaystyle 1= \frac{2 \rho_0 ( \sigma_0^2 + x_0^2)}{- y_F} }\ ,  \\
{\displaystyle \mu = - \frac{(\sigma_0^2 + x_0^2)}{D} }\ ,   
\end{array} \ee where $y_{F}=\mu-\ef$ and $\rho_0 = 1/2D$ is the bare density of states of conduction electrons. From this set of equations, we find : $\ef =0$.

\begin{figure}
\centerline{\psfig{file=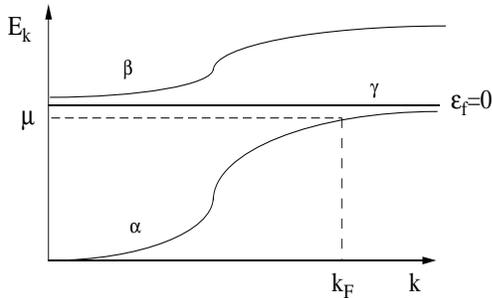,height=4cm,width=7cm}}
\caption{Sketch of energy versus wave number $k$ for the three bands $\alpha$, $\beta$, $\gamma$ resulting of the diagonalization of supersymmetric $H_0$. Note the presence of a bosonic band $\gamma$ separating the fermionic bands $\alpha$ and $\beta$. Note also the flatness of the band at $k=k_F$ reflecting large effective mass.}
\label{fig1}
\end{figure}

The resulting spectrum of energies is schematized in Figure~\ref{fig1}.
At zero temperature, only the lowest band $\alpha$ is filled with an enhancement of the density of states at the Fermi level (and hence of the mass) unchanged from the standard $1/N$ expansions: $ {\displaystyle \frac{\rho(E_F)}{\rho_0} = 1+ \frac{(\sigma_0^2 + x_0^2)}{y_F^2} = 1+ \frac{D}{(-y_F)} \gg 1 }$. This large mass enhancement is related to the flat part of the $\alpha$ band associated with the formation of the Abrikosov-Suhl resonance pinned 
at the Fermi level. While this feature was already present in the purely fermionic description, it is to be noted that the formation of a dispersionless bosonic band within the hybridization gap is an entirely new result of the theory.

\begin{figure}
\centerline{\psfig{file=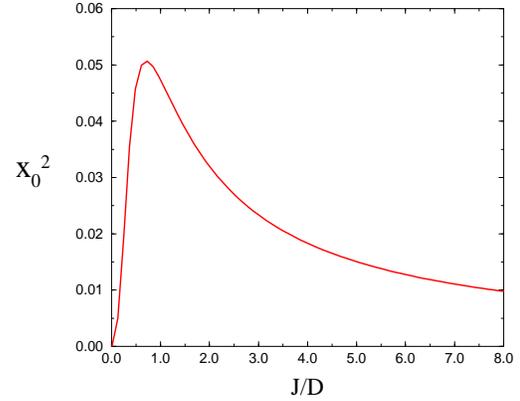,height=6cm,width=7cm}}
\caption{$J/D$-dependence of the coupling $x_0^2= \protect \left\langle \eta \eta^* \protect \right \rangle$ fixing the relative weight of fermion and boson statistics. The unit on the vertical-axis is $D^2$.}
\label{fig2}
\end{figure}

The relative weight of boson and fermion statistics in the spin representation is related to $x_0^2$ : $n_b/n_f=x_0^2/{\sigma_0^2}$. It is then interesting to
follow the J-dependence of $x_0^2$. The result is reported in Fig.~\ref{fig2}. This bell-shaped curve can be interpreted in the light of the exhaustion principle mentioned in the introduction. In the
limit of large J, the Kondo temperature-scale $T_K=D exp[-1/(2J\rho_0)]$ is of order of the bandwidth. One then expects a complete Kondo screening as can be checked by remarking that the weight of c in the $\alpha$ quasiparticle at the Fermi level (noted $v_{k_F}^2$) just equals the added weights of f and b at the Fermi level (respectively noted $u_{k_F}^2$ and $\rho_1^2$): $v_{k_F}^2/(u_{k_F}^2+\rho_1^2)=y_F^2/(\sigma_0^2+x_0^2)=1$. The Kondo effect being complete in that limit, there is no residual unscreened spin degrees of freedom: it is then natural to derive a zero value of $x_0^2$ (and hence of $n_b$). The opposite limit at small J corresponds to the free case of uncoupled impurity spins and conduction electrons. It also leads to: $x_0^2=0$. The finite value of $x_0^2$ between these two limits with a maximum reflects the incomplete Kondo screening effect in the Kondo lattice, the unscreened spin degrees of freedom being described by bosons. 

Largely discussed in the litterature   \cite{ueda}  is the question concerning the Fermi surface sum rule: do the localized spins of the Kondo lattice contribute to the counting of states within the Fermi surface or do they not? Depending on the answer, one expects large or small Fermi surfaces. The supersymmetric theory leads to a firm conclusion in favour of the former. One can check that the number of states within the Fermi surface is just equal to $n_c+n_b+n_f$, i.e. $n_c+1$. The Fermi surface volume includes a contribution of one atate per localized spin in addition to that of conduction electrons \cite{lonzarich}. This conclusion which appears reasonable if one recalls that the KLM is an effective hamiltonian derived from the PAM, has been reached before by several other authors \cite{ueda,martin}. We do think that it is a good sign to recover previously established results if they are correct.

The dynamic spin susceptibility $\chi({\bf q}, \omega)$
is calculated as usual from the second derivative of the generating function $lnZ(H_i)$ with respect to a magnetic source $H_i$. The corresponding spin-spin correlation function is given by: $<(S_f(t)+S_b(t))(S_f(0)+S_b(0))>$. Projecting it in the eigenbasis $\{\alpha, \beta, \gamma \}$ and omitting negligible terms, one gets two contributions to $\chi"({\bf q}, \omega)$ : 
\be
\chi"({\bf q}, \omega)
=\chi"_{\alpha\alpha}({\bf q}, \omega)+\chi"_{\alpha\gamma}({\bf q}, \omega)
\ee
The first one is of Lindhard type and corresponds to electron-hole pair excitations in the $\alpha$ band: 
\be
\chi"_{\alpha\alpha}({\bf q}, \omega) \sim \rho(E_F) \frac{\omega}{qv_F^*}
\ee 
where $v_F^*$ is the renormalized Fermi velocity.
The second one corresponding to the excitations from the $\alpha$ to the $\gamma$ band has a peak at a frequency of order $T_K$:
\be
\chi"_{\alpha\gamma}({\bf q}, \omega) \sim \rho_0 \frac{1}{\omega^2(\sigma_{0}^{-2}+x_{0}^{-2})} \theta(\omega-y_F)
\ee
We believe that these two contributions
may account for the quasielastic and inelatic peaks observed in INS.

\begin{figure}
\centerline{\psfig{file=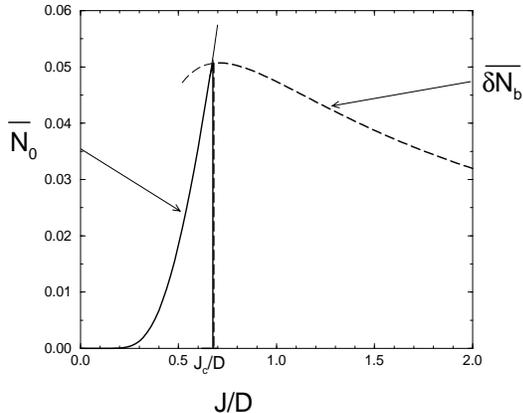,height=6cm,width=7cm}}
\caption{$J/D$-dependence of the renormalized populations ${\overline N_0}$ and ${\overline {\delta N_b}}$ (where ${\overline N_0}=N_0.T_K$ and  ${\overline {\delta N_b}}=\delta N_b.T_K$) of the bosons in the condensate and the excited levels respectively. The intersection of the two curves determine the critical value $J_c/D=0.67$ below which long-range ferromagnetic order appears. Note the first order transition at $J_c/D$ with ${\overline N_0}$ (and ${\overline {\delta N_b}}$) jumping to (and from) zero at this value.}
\label{fig3}
\end{figure}

One of the crucial questions raised by the Kondo lattice model is the possibility of long-range magnetic order at low $J$. So far, all the results
have been obtained assuming that no Bose condensation occurs. Let us now reconsider the problem allowing the bosons b condense. At the level of approximation that we use, bosons behave as free particles; so at $T=0$, the bosons are either all in the excited levels or all in the condensate. In the latter case, we evaluate the number of bosons $N_0$ in the condensate to $N_0 = \exp[-1/2 \rho_0 J]$. This value should be put together with the number of bosons $ \delta N_b$ in the excited levels in the absence of Bose condensation: $ \delta N_b=2\rho_0 x_0^2/(-y_F) = x_0^2/(D^2 \exp[-1/2 \rho_0 J]) $. It turns out that comparing the free energy of both states is equivalent to compare $N_0$ and $\delta N_b$. Figure~\ref{fig3} represents the $J$-dependence of the two populations. The intersection of the two curves defines the value of $J$ below which the bosons condense. The ferromagnetic instability associated to this Bose condensation appears at $J_c/D = 0.67$. This value is significantly reduced compared to the fermionic theory mean-field value of order $1.0$ \cite{lacroix,read}. The reduction of $J_c/D$ that we predicted is similar to what is found in the related problem of the 1D Kondo necklace \cite{ueda} for which numerical calculations also lead to smaller critical value ($J_c/D =0.382$ and $0.24$ within real-space renormalization group and finite size scaling analysis respectively) compared to mean-field value ($J_c/D = 1.0$). It is obvious that the sharp transition of $N_0$ and $\delta N_b$ at $J_c$ associated with a first-order transition is an artifact of our approximations. Further refinements including fluctuations would introduce interactions among bosons, as in the Bogoliubov problem. We expect that it will produce progressive depletion of the condensate towards the excited levels transforming the ferromagnetic transition from first to second order. With a few exceptions, most of the heavy Fermions are not ferromagnetic but rather antiferromagnetic. The present calculation cannot directly apply. However, we believe that the discussion here is very useful since (i) it draws the general lines of the study which could as well be extended to the case of antiferromagnetism by considering a staggered instead of an uniform condensate and (ii) it can be considered as a test of the theory by comparing the prediction of the ferromagnetic instability with the results of numerical calculations.

In summary, we have shown that the supersymmetric method proves to be a powerful tool to account for both Fermi-liquid excitations and residual spin degrees of freedom in heavy-Fermion systems. The key results are the prediction of a magnetic instability at $J_c/D=0.67$ significantly reduced compared to purely fermionic theories, the presence of two contributions in the dynamical spin susceptibility, the existence of an enhanced density of states at the Fermi level while the Fermi surface sum rule includes the total number of $n_c+1$ states. The work opens the way for further investigations as the systematic study of the effects of fluctuations \cite{fluct} in the pure Kondo lattice model or the incorporation of additional RKKY interactions in an extended Kondo lattice model.

We would like to thank P.Coleman, A.Georges, P.A.Lee, A.Yashenkin and T.Ziman for very helpful discussions.

* Also Part of the Centre National de la Recherche Scientifique (CNRS).

\par
	\vspace{-.5  cm} 



\begin{references}
	\vspace{-1.2 cm} 

\bibitem{aeppli} G. Aeppli, H. Yoshizawa, Y. Endoh, E.Bucher, J. Hugnagl, Y. Onuki, T.Komatsubara, Phys.Rev.Lett. {\bf 57}, 122 (1986). 
\bibitem{rossat} L.P. Regnault, W.A.C. Erkelens, J. Rossat-Mignod, R. Lejay, J. Flouquet, Phys.Rev.B {\bf 38}, 4481 (1988). 
\bibitem{lonzarich} S.R. Julian, F.S. Tautz, G.J. McMullan, G.G. Lonzarich, PhysicaB {\bf 199 \& 200}, 63 (1994). 
\bibitem{nozieres} P. Nozi\`eres, Ann.Phys.Fr. {\bf 10}, 19 (1985). 
\bibitem{millis} A.J. Millis, P.A. Lee, Phys.Rev.B {\bf 35}, 3394 (1987). 
\bibitem{auerbach} A. Auerbach, K.Levin, Phys.Rev.Lett. {\bf 57}, 877 (1986). 
\bibitem{efetov} K.B.Efetov, Adv. Phys. {\bf 32}, 53 (1983). 
\bibitem{gan} J.Gan, P.Coleman, N.Andrei, Phys.Rev.Lett. {\bf 68}, 3476 (1992). 
\bibitem{ref3} We extend here the equation used in Ref.[7] for the determination of eigenenergies for fermions to the case of bosons.  
\bibitem{doniach} S. Doniach, Physica {\bf 91B}, 231 (1977). 
\bibitem{ueda} H. Tsunetsugu, M. Sigrist, K. Ueda (in press) and references within. 
\bibitem{martin} R.M. Martin, Phys.Rev.Lett. {\bf 48}, 362 (1982). 
\bibitem{lacroix} C. Lacroix, M.Cyrot, Phys.Rev.B {\bf 20}, 1969 (1979). 
\bibitem{read} N. Read, D.M. Newns, S. Doniach, Phys.Rev.B {\bf 30}, 3841 (1984). 
\bibitem{fluct} Note that part of the fluctuation effects has already been taken into account in the local approximation. 
\end{references}
\end{document}